\documentclass{elsart}

\usepackage{amssymb}

\begin{document}

\begin{frontmatter}



\title{Quantum Auctions: Facts and Myths\thanksref{grant}}
\thanks[grant]{This research was supported in part by the {\bf Polish Ministry of Science
 and Higher Education project No N519 012 31/1957}.}

\author[eda]{Edward W. Piotrowski}

\address[eda]{Institute of Mathematics,
University of Bia\l ystok, Lipowa 41, Pl 15424 Bia\l ystok,
Poland}\ead{ep@wf.pl}
\author[jaa]{Jan S\l adkowski\corauthref{cor1}}
\address[jaa]{Institute of Physics, University of Silesia, Uniwersytecka
4, Pl 40007 Katowice, Poland}\ead{jan.sladkowski@us.edu.pl }
\corauth[cor1]{Corresponding author.}

\begin{abstract}
Quantum game theory, whatever opinions may be held due to its
abstract physical formalism, have already found
 various applications even outside the orthodox physics domain.
 In this paper we  introduce the concept of a quantum auction, its advantages and drawbacks. Then
 we  describe the models that have already been put forward. A general model involves
 Wigner formalism and infinite dimensional Hilbert spaces -- we
 envisage that the implementation
 might not be an easy task.  But a restricted model advocated by the Hewlett-Packard
 group seems to be much easier to implement. Simulations involving humans have
 already been performed. We will focus on
 problems related to combinatorial auctions and technical assumptions that are
 made. Quantum approach offers at least two important
 developments.  Powerful quantum algorithms for finding solutions would extend the range of
 possible applications. Quantum strategies, being qubits, can be teleported but
 are immune from cloning -- therefore extreme privacy of agent's activity could in principle be guaranteed.
 Then we point out some key problem that have to
 be solved before commercial use would be possible. With present technology, optical networks, single photon sources and detectors
 seems to be sufficient for experimental realization in the near future. We  conclude by describing potential
 customers,
 estimating the potential market size and possible timing.
\end{abstract}

\begin{keyword}
quantum game\sep quantum information theory\sep quantum
markets\sep quantum finance\sep auction theory
\PACS  02.50.Le\sep 03.67.Lx\sep 05.50.+q\sep 05.30.–d
\end{keyword}
\end{frontmatter}

% main text
\vspace{5mm}
\begin{flushleft}
{\bf Motto:}
\end{flushleft}

\begin{center}
{\em "Combinatorial auctions are the great frontier of auction
theory today,
. . . ". }\\
   \,\hfill
   \scriptsize Roger B. Myerson on the back cover of Peter Cramton, Yoav Shoham and Richard Steinberg (Eds.), Combinatorial Auctions,
MIT Press, Cambridge, 2006.\\
\end{center}

\vspace{5mm}
\section{Introduction}
%\label{}
 Quantum game theory \cite{Mey} - \cite{inv} emerged  as an abstract
idea in quantum theory but soon it was realized that it offers
powerful analytical tools that might be used outside physical
laboratories! Game theory, the study of (rational) decision making
in conflict situations, seems to have asked for a quantum version.
For example, games against Nature  include those for which Nature
is quantum mechanical. But does quantum theory offer more subtle
ways of playing games? Game theory considers strategies that are
probabilistic mixtures of pure strategies. Why cannot they be
intertwined in a more complicated way, for example interfered or
entangled? The research already performed suggests that there are
several possible niches\footnote{ The point is that niches should
not be equated with small but rather you should think of narrow:
the targeting at a more narrowly defined customer group seeking a
distinctive source of benefits. Niche markets are not the marginal
opportunity that they  once used to be.} for the quantum game
products launch. The most promising seem to be quantum
cryptography, "quantum" hazard and quantum auctions. One can
already buy quantum cryptographic equipment provided by {\it id
Quantique, MagiQ Technologies, SmartQuantum} and, what is more
important, many well known industrial concerns have revealed their
their interest in quantum technologies, not to mention
military/security oriented projects. Although the market may be
worth of billions dollar the involved complication obstruct
massive application. This would certainly change if the security
of the presently used cryptographic systems is challenged by the
increase in computational power or other developments. Quantum
hazard has big potential and it seems that present technology is
sufficient for implementation. The implementation will be costly,
but if you compare the estimated costs of the order of $\$10^8$
with
 the amount of money spent on advertising related products the situation seems to be
 promising! Optical cluster states  presently form the most promising implementation environment \cite{cluster1, cluster2}.
 The key issue is to invent a simple to implement, possibly interesting (drawing in), quantum game
 -- the inventor would get the due gratification! The first, to our knowledge, proposal was put forward
 in ref. \cite{casino}. Although it is
implementable, one can hardly say it would be exciting for
non-physicists. On the other hand, quantum auctions, if ever
implemented, would be designed for very specific and limited
business circles: the volume must be huge and and items combined.
The paper is organized as follows. We will begin by presenting the
general idea of a quantum game and methods of gaining an advantage
over "classical opponent". Then we will attempt at giving a
definition of a quantum auction and review problems that have
already been discussed in the literature. Finally we will try to
show some problems that should be addressed in the near future. In
the following discussion we will use quantum auction theory as a
formal theoretical tool but the broadcasted message would be that
it would probably be used massive combinatorial auctions in the
future \cite{patel} or in compound securities trading
\cite{Fortnow}.
\section{Quantum games}
It is not easy to give the precise date of birth of quantum game
theory. Quantum games have  been with us camouflaged since the
very beginning of the  quantum era because a lot of experiments
can be reformulated in terms of game theory. Quantum game theory
began with works of Wiesner on quantum money \cite{Wie}, Vaidman,
who probably first used the term  game in quantum context
\cite{Vai} and Meyer \cite{Mey}  and Eisert et al \cite{Eis} who
first formulated their problems in game theory formalism. Possible
applications of quantum games in biology are thoroughly discussed
by Iqbal  \cite{Iqbal}, in economics by Piotrowski and S\l
adkowski \cite{PS1,qf}. Flitney and Abbott quantized Parrondo's
paradox \cite{Abb}. The most popular experimental realizations are
described in refs \cite{Du1, cluster2}. In principle, any quantum
system that can be manipulated  by at least one party and where
the utility of the moves can be reasonably defined, quantified and
ordered may be conceived as a quantum game\footnote{One can also
consider the class of games against Nature.}. The quantum system
may be referred to as a {\it quantum board}\/ although the term
{\it universum of the game}\/ seems to be more appropriate
\cite{Bug}. Usually one  supposes that all players know the state
of the game at the beginning and at some crucial stages that may
depend an the game being played. This is a subtle point because it
is not always possible to identify the state of a quantum system
let alone the technical problems with actual identification of the
state (one can easily give examples of systems that are only
partially accessible to some players \cite{Bug2}). A "realistic"
quantum game should include measuring apparatuses or information
channels that provide information on the state of the game at
crucial stages and specify the way of its termination. Therefore
we will suppose that a {\it two--player quantum game}\/
$\Gamma\negthinspace =\negthinspace({\cal
H},\rho,S_A,S_B,P_A,P_B)$ is completely specified by the
underlying Hilbert space ${\cal H}$ of the physical system, the
initial state $\rho\negthinspace\in\negthinspace {\cal S}({\cal
H})$, where ${\cal S}({\cal H})$ is the associated state space,
the sets $S_A$ and $S_B$ of permissible quantum operations of the
two players, and the { pay--off (utility) functions}\/ $P_A$ and
$P_{B\/}$, which specify the pay--off for each player. A {\it
quantum strategy}\/ $s_A\negthinspace\in\negthinspace S_A$,
$s_B\negthinspace\in\negthinspace S_B$ is a collection of
admissible quantum operations, that is the mappings of the space
of states onto itself. One usually supposes that they are
completely positive trace preserving maps. The quantum game's
definition may also include certain additional rules, such as the
order of the implementation of the respective quantum strategies
or restriction on the admissible communication channels, methods
of stopping the game etc. The generalization for the N players
case is obvious.
 Schematically we have:
$$
\rho \mapsto (s_{A},s_{B},\ldots) \mapsto  \sigma \Rightarrow
(P_{A}, P_{B},\ldots)\, ,$$ where $\sigma$ denote the measurement
of state of the game combined with the prize allocation algorithm.
\section{Quantum auctions}
Quantum auction are quantum games designed for goods allocations.
Some of the researchers involved believe that they might be some
day an alternative for "classical" auctions designed in cases
where combinatorial and computational problems hinder the
designers in their work. Currently, it is difficult to find out if
this is a feasible task. Bellow, we describe some proposals that
have already been put forward. Shortly, a protocol for a quantum
auction should specify the following steps.
\begin{itemize}
    \item
   Auctioneer specifies
conventional "classical" details of the auction such as the
schedule, goods to be sold etc. \item Auctioneer specifies the
implementation of the quantum auction.
  \item Auctioneer specifies the initial state distribution, implementation of strategies
and main features of the search algorithms to be used (eg
probabilistic, deterministic etc).
 \item Search for the winner{s} and good allocations (this process might
be repeated several times). \item Methods of good delivery and
clearing.
\end{itemize}
The first and the last items are not directly connected with the
"quantumness" of the auction and will not be discussed here.
Schematically we can write
$$ \rho \mapsto (s_{1},s_{2},\cdots,s_{n}) \mapsto  \sigma
\Rightarrow (P_{1},\cdots ,P_{n})\, ,$$ where $s_i$ and $P_i$
denote bids goods allocation, respectively.
\subsection{GG model} An interesting model of, roughly speaking, a
population of N quantum bargaining games being played on
  a market was recently proposed by Gon\c{c}alves and
  Gon\c{c}alves \cite{ISCTE}. The idea behind it is that one can
  introduce a "population numbers" $n_1,n_2,\ldots,n_m$ for all
  alternative strategies combinations. This fact is described by
  bosonic creation and annihilation operators $a^{\dag}_{k}$ and
  $a_{k}$with standard commutation relation. The number of all
  possible combination, $m=\prod _k N_k$ is unlimited ($N_j$ is
  the number of alternative strategies for the $j$-th player.
In general, the $j$-th agent strategy profile is $|
p_{j}\rangle=\sum _{i}c_{i} |s_{i}( p_{j})\rangle$, where $c_i$ is
the probability amplitude of strategy $s_i$ The unitary evolution
of the strategy state  $| p_{j}, t_{fin}\rangle
=U(t_{fin},t_{ini}) |p_{j},t_{ini}\rangle\negthinspace$ is
governed by a unitary operator of the form
$$U(t_{fin},t_{ini})=\prod _{k=0}^{k_{fin}}U(t_{k+1},t_{k}), $$
where $k$ parameterizes the $k_{fin}+1$ trading rounds. In a
simplified single-asset model, where there are only two strategies
(buying and selling) for each agent the state $| n_0, n_1\rangle$
is characterized by two occupation numbers $n_0$ and $n_1$ giving
the number of agents that are selling and buying, respectively.
Then the unitary evolution for the $k$-th trading round can be
given in the following form:
  $$U(t_{k+1},t_{k})=
\exp (\sum _{j=0}^{1}(\xi _{j}(k,\tau _{k})a_{j}^{\dagger}
  -\xi _{j}(k,\tau _{k})^{\ast}a_{j})), $$ where $\tau _{k}$ is
  the duration of each trading round. Even this oversimplified
  model reproduces multifractal signatures similar to those of
  real markets \cite{ISCTE}. This is an interesting analytical tool -- the model can be run online at the
web\footnote{
http://ccl.northwestern.edu/netlogo/models/community/Quantum\_Financial\_Market}.
Experimental implementation would not be easy, but  the mastering
of coherent photon states might render it feasible.
\subsection{PS model}
 Piotrowski and S\l adkowski \cite{PS1} put forward a quantum model of bargaining in infinite
 dimensional Hilbert space. The $k$-th agent strategies $|
p_{j}\rangle _k$ belong to Hilbert spaces $H_k$. The initial state
of the game $|\Psi\rangle_{in}:=\sum_k|\psi \rangle_k$ is a vector
in the direct sum of Hilbert spaces of all players, $\oplus _{k}
H_k$. Then one defines canonically conjugate hermitian operators
of demand $\mathcal{Q}_k$ and supply $\mathcal{P}_k$ for each
Hilbert space $H_k$ analogously to their physical counterparts,
position and momentum. The observable
$$
H(\mathcal{P}_k,\mathcal{Q}_k):=\frac{(\mathcal{P}_k-p_{k0})^2}{2m}+
                     \frac{m\omega^2(\mathcal{Q}_k-q_{k0})^2}{2},
                     \eqno(12)
$$
where $p_{k0}:=\frac{
\phantom{}_k\negthinspace\langle\psi|\mathcal{P}_k|\psi\rangle_k }
{\phantom{}_k\negthinspace\langle\psi|\psi\rangle_k}$ $\neq
E(\mathcal{P}_k)$, $q_{k0}:=\frac{
\phantom{}_k\negthinspace\langle\psi|\mathcal{Q}_k|\psi\rangle_k }
{\phantom{}_k\negthinspace\langle\psi|\psi\rangle_k}$,
$\omega:=\frac{2\pi}{\theta}$, and is called {\it the risk
inclination operator}, cf Ref. \cite{acta}. $ \theta$ denotes the
characteristic time of transaction introduced in the MM model
\cite{3}. Noncommuting variables appear in a natural way here
\cite{acta}. A transaction consists in a transition from the state
of traders strategies $|\Psi\rangle_{in}$ to the one describing
the capital flow state $|\Psi\rangle_{out}:=\mathcal{T}_\sigma
|\Psi\rangle_ {in}$, where
$\mathcal{T}_{\sigma}:=\sum_{k_d}|q\rangle_{k_d}\phantom{}_{k_d}
   \negthinspace\langle q|+
 \sum_{k_s}|p\rangle_{k_s}\phantom{}_{k_s}
   \negthinspace\langle p|$  is the projective operator defined by
the division $\sigma $ of the set of traders $\{ k\}$ into two
separate subsets $\{k\}=\{k_d\}\cup\{k_s\}$, the ones buying at
the price $e^{q_{k_d}}$ and the ones selling at the price
$e^{-p_{k_s}}$ in the round of transactions in question. The key
role is played by a (quantum) algorithm
 $\mathcal{A}$ that determines the division
$\sigma$ of the market, the set of price parameters $\{ q_{k_{d}},
p_{k_{s}}\}$ and the values of capital flows. The capital flows
resulting from an ensemble of simultaneous transactions correspond
to the physical process of measurement. The later are settled by
the distribution $$\int_{-\infty}^{\ln c} \frac{{|\langle
q|\psi\rangle_k|}^2}{\phantom{}_k
\negthinspace\langle\psi|\psi\rangle_k}dq $$ which is interpreted
as the probability that the trader $| \psi \rangle _{k}$ is
willing to buy  the asset $\mathfrak{G}$ at the transaction price
$c$ or lower.
 In an analogous way the distribution
$$ \int_{-\infty}^{\ln \frac{1}{c}} \frac{{|\langle
p|\psi\rangle_k|}^2}{\phantom{}_k
\negthinspace\langle\psi|\psi\rangle_k}dp  $$ gives the
probability of selling $\mathfrak{G}$ by the trader $| \psi
\rangle _{k}$ at the price $c$ or greater.

 These probabilities
are in fact conditional because they describe the situation after
the division $ \sigma $ is completed. Various possible class of
tactics and strategies are discussed in Refs \cite{PS1}.
Experimental implementation would be hard but quantum markets
would have such astonishing features the that it is worth a try!

\subsection{The HP group model} In this model any possible price of each item (multiply auctions are possible)
 are encoded in strings of qubits \cite{HP1}-\cite{HP2}. The bidder
 specifies the his bid by selecting the corresponding vector of the Hilbert space --
 each bidder gets p, $p=p_{item} + p_{price}$
qubits and can only operate on those bits. Thus each bidder has
$2^p$ possible bids values, and can create superpositions of these
bids: for multiply item auction the bid is a superposition $\sum
_{j} \alpha _{j} |bundle _{j}\rangle \otimes |price_{j}\rangle$
for each bundle of items. A superposition of bids specifies set of
distinct bids, with at most one allowed to win and amplitudes of
the superposition correspond to the likelihood of various outcomes
for the auction. The protocol uses a distributed adiabatic search
that guarantee that bidder's strategies remain private \cite{HP3}.
The search operation processing input from the bidders implemented
by unitary operators, giving the overall operator $U = U_1\otimes
U_2\otimes \ldots \otimes U_n (1)$, where $n$ is the number of
bidders and $U_i$ the operator of $i$-th bidder \cite{HP1}. A
perfect search is not always possible and probabilistic outcomes
should be allowed for. Some additional sub-procedures might be
necessary to prevent dishonest agents and auctioneers from getting
advantage \cite{HP2}. In such a "brute force" implementation the
existence of equilibria can be proved. This proposal seems to be
the easiest to implement and especially suitable for combinatorial
 auctions. Details and simulations are given in Refs
 \cite{HP1}-\cite{HP4}.

\section{Conclusion}
Quantum auctions are certainly an interesting theoretical
alternative for complex and massive auctions but
 are they feasible? Encoding bids in quantum states is a challenge to (quantum) game
theory: quantum auctions would almost always be probabilistic and
may provide us with specific incentive mechanisms etc.  As the
outcome may depend on amplitudes of quantum strategies
sophisticated apparatus and specialist (physicists?) may be
necessary. Therefore, we envisage some changes in the law and
habits. Combinatorial auctions seem to be the most promising
field. But despite the promising quantum-like experiments
\cite{HP4} commercial implementation of quantum auctions  is a
demanding challenge that would hardly be accomplished without a
major technological breakthrough in mastering quantum devices.
Recent development in quantum information processing raises many
important issues \cite{raport}: {\it Are markets predestined to
quantum technologies? Would we rise to the challenge?} For the
present, quantum game theory is only and interesting theoretical
tool in various fields of research but the situation might soon
change in a dramatic way!

% \appendix

% \section{}
% \label{}

\end{document}